\documentstyle[12pt]{article}

\newcommand{\ejp}[4]{{\em #1} - Eur.~J.~Phys.           {\bf #2},  #3,  (19#4)}

\newcommand{\nc}[4]{{\em #1} - Nuovo Cim.               {\bf #2},  #3,  (19#4)}

\newcommand{\pr}[4]{{\em #1} - Phys.~Rev.               {\bf #2},  #3,  (19#4)}

\newcommand{\xx}[3]{                                    {\bf #1},  #2,  (19#3)}

\newcommand{\bel}[1]{\begin{equation}\label{#1}}
\newcommand{\be}{\begin{equation}}
\newcommand{\ee}{\end{equation}}

\newcommand{\beal}[1]{\begin{eqnarray}\label{#1}}
\newcommand{\bea}{\begin{eqnarray}}
\newcommand{\eea}{\end{eqnarray}}

\newcommand{\bean}{\begin{eqnarray*}}
\newcommand{\eean}{\end{eqnarray*}}

\newcommand{\ba}{\begin{array}}
\newcommand{\ea}{\end{array}}

\newcommand{\bamq}[4]{\left( \begin{array}{cccc}{#1}&{#2}&{#3}&{#4}\\}
\newcommand{\bamc}[5]{\left( \begin{array}{ccccc}{#1}&{#2}&{#3}&{#4}&{#5}\\}
\newcommand{\eam}{\end{array} \right)}

\newcommand{\Raw}{\Rightarrow}

\newcommand{\ag}{\alpha}
\newcommand{\bg}{\beta}
\newcommand{\cg}{\gamma}
\newcommand{\dg}{\delta}
\newcommand{\eg}{\epsilon}

\newcommand{\tg}{\theta}

\newcommand{\vpg}{\varphi}
\newcommand{\og}{\omega}

\newcommand{\Lg}{\Lambda}

\newcommand{\Og}{\Omega}

\newcommand{\Lb}{\Large \bf}

\newcommand{\fs}{\footnotesize}

\newcommand{\bii}{\begin{itemize}}
\newcommand{\eii}{\end{itemize}}

\newcommand{\ben}{\begin{enumerate}}
\newcommand{\een}{\end{enumerate}}

\newcommand{\bq}{\begin{quote}}
\newcommand{\eq}{\end{quote}}

\newcommand{\bc}{\begin{center}}
\newcommand{\ec}{\end{center}}

\newcommand{\bt}{\begin{tabular}}
\newcommand{\et}{\end{tabular}}

\newcommand{\br}{\begin{flushright}}
\newcommand{\er}{\end{flushright}}

\newcommand{\bl}{\begin{flushleft}}
\newcommand{\el}{\end{flushleft}}

\newcommand{\f}[1]{\footnote{#1}}

\newcommand{\vs}[1]{\vspace*{#1}}

\newcommand{\new}{\pagebreak}

\newcommand{\bb}{\begin{thebibliography}{99}}
\newcommand{\eb}{\end{thebibliography}}
\newcommand{\bi}{\bibitem}

\newcommand{\btp}{\begin{titlepage}}
\newcommand{\etp}{\end{titlepage}}

\newcommand{\go}{\section{Introduction}}

\begin{document}

\hyphenation{transfor-ma-tion introdu-ce  Lo-ren-tz qua-ter-ni-on
transformati-ons manipulati-ons com-ple-te intro-du-ce com-plexi-fi-ed
Di-rac in-ter-national}

\btp

\bc
August, 1995 \hfill {\fs hep-th/95mmnnn}

\vs{3cm}

{\Lb Quaternions and Special Relativity}

\vs{2cm}

{\sc Stefano De Leo}$^{ \; a)}$\\

\vs{1cm}

{\it Universit\`a  di Lecce, Dipartimento di Fisica\\
Instituto di Fisica Nucleare, Sezione di Lecce\\
Lecce, 73100, ITALY}

\vs{2cm}

{\bf Abstract}

\ec
We reformulate Special Relativity by a quaternionic algebra on reals.
Using {\em real linear quaternions}, we show that previous difficulties,
concerning the appropriate transformations on the $3+1$ space-time, may be
overcome. This implies that a complexified quaternionic version of
Special Relativity is a choice and not a necessity.

\vs{2cm}

\noindent{\fs a) e-mail:} {\fs \sl deleos@le.infn.it}

\etp

\new

\go

``{\sl The most remarkable formula in mathematics is:
\begin{equation}
e^{i\theta} = \cos{\theta} + i \sin{\theta} \; \; .
\end{equation}
This is our jewel. We may relate the geometry to the algebra by
representing complex numbers in a plane
\[ x+iy = re^{i\theta} \; \; . \]
This is the unification of algebra and geometry.}'' -
Feynman~\cite{fey}.\\

We know that a rotation of $\alpha$-angle around the $z$-axis, can be
represented by $e^{i\alpha}$, in fact
\[ e^{i\ag}(x+iy) = re^{i(\theta + \ag)} \; \; . \]
In 1843, Hamilton in the attempt to generalize the complex
field in order to describe the rotation
in the three-dimensional space, discovered quaternions\f{Quaternions, as
used in this paper, will always mean ``real quaternions''
\[q=a+ib+jc+kd \; \; \; \; , \; \; \; \;
a, \; b, \; c, \; d \; \in \; {\cal R} \; \; .\]}.\\
Today a rotation about an axis passing trough the origin and parallel to a
given unitary vector $\vec{u} \equiv (u_{x}, \; u_{y}, \; u_{z})$ by an
angle $\ag$, can be obtained taking the transformation
\be
e^{(iu_{x}+ju_{y}+ku_{z})\frac{\ag}{2}} (ix+jy+kz)
e^{-(iu_{x}+ju_{y}+ku_{z})\frac{\ag}{2}} \; \; .
\ee
Therefore if we wish to represent rotations in the three-dimensional space
and complete ``{\sl the unification of algebra and geometry}'' we need
quaternions.

The quaternionic algebra has been expounded in a series of papers~\cite{pap}
and books~\cite{boo} with particular reference to quantum mechanics, the reader
may refer to these for further details. For convenience we repeat and
develop the relevant points in the following section, where the terminology
is also defined.

Nothing that $U(1, \; q)$ is algebraically isomorphic to $SU(2, \; c)$, the
imaginary units $i, \; j, \; k$ can be realized by means of the
$2\times 2$ Pauli matrices through
\[ (i, \; j, \; k) \leftrightarrow (i\sigma_{3}, \; -i\sigma_{2}, \;
-i\sigma_{1}) \; \; . \]
\bc
{\fs (this particular representation of the imaginary units $i, \; j, \; k$
has been introduced in ref.~\cite{del}).}
\ec
So a quaternion $q$ can be represented by a $2\times 2$ complex matrix
\bel{a}
q \; \; \leftrightarrow \; \; {\cal Q} = \left(
\begin{array}{cc} z_{1} & -z_{2}^{*}\\
z_{2} & z_{1}^{*}\end{array} \right) \; \; ,
\ee
where
\[ z_{1}=a+ib \; \; \; , \; \; \; z_{2}=c-id \; \; \; \; \in \; \;
{\cal C}(1, \; i) \; \; .\]
It follows that a quaternion with unitary norm is identified by a
unitary $2\times 2$ matrix with unit determinant, this gives the
correspondence between unitary quaternions\f{In a recent
paper~\cite{del2} the representation theory of the group $U(1, \; q)$ has been
discussed in detail.} $U(1, \; q)$ and $SU(2, \; c)$.\\
Let us consider the transformation law of a spinor (two-dimensional
representations of the rotation group)
\be
{\psi}'={\cal U} \psi \; \; ,
\ee
where
\[ \psi=\left( \begin{array}{c} z_{1}\\ z_{2}\end{array} \right) \; \;
\; \; , \; \; \; \; {\cal U} \in SU(2, \; c) \; \; . \]
We can immediately verify that
\[ \tilde{\psi}=\left( \begin{array}{c} -z_{2}\\ z_{1}\end{array} \right) \]
transforms as follows
\be
{\tilde{\psi}}'={\cal U}^{*}\tilde{\psi} \; \; ,
\ee
so
\[\left( \begin{array}{cc} z_{1} & -z_{2}^{*}\\
z_{2} & z_{1}^{*}\end{array} \right)' = {\cal U}
\left( \begin{array}{cc} z_{1} & -z_{2}^{*}\\
z_{2} & z_{1}^{*}\end{array} \right) \]
represents again the transformation law of a spinor.\\
Thanks to the identification~(\ref{a}) we can write the previous
transformations by real quaternions as follows
\[ q'={\cal U}q \; \; , \]
with $q=z_{1}+jz{_2}$ and $\cal U$ quaternion with unitary norm
{\fs ($N({\cal U})={\cal U}^{+}{\cal U}=1$)}. Note
that we don't need right operators to indicate the transformation law of a
spinor.

Now we can obtain the transformation law of a three-dimensional
vector $\vec{r}\equiv (x, \; y, \; z)$ by product of spinors, in fact
if we consider the purely imaginary quaternion
\[ \og=qiq^{+}=ix+jy+kz \]
or the corresponding traceless $2\times 2$ complex matrix
\[ \Og =\psi i \psi^{+} = \left( \begin{array}{cc} ix & -y-iz\\
y-iz & -ix\end{array} \right) \; \; , \]
a rotation in the three-dimensional space can be written as
follows\f{``{\sl No such `trick' works to relate the full
four-vector $(ct, \; x, \; y, \; z)$ with real
quaternions.}'' - Penrose~\cite{pen}.}
\bc
\bt{cl}
${\og}'={\cal U}\og \; {\cal U}^{+}$ &
{\fs (quaternions)} ,\\ \\
${\Og}'={\cal U}\Og \; {\cal U}^{+}$ & {\fs ($2\times 2$ complex matrices)} .
\et
\ec
For infinitesimal transformations, ${\cal U}=1+\vec{Q}\cdot \vec{\tg}$,
we find
\[ \vec{Q}\cdot \vec{r} \; ' = \vec{Q}\cdot \vec{r}+
\vec{Q}\cdot \vec{\tg} \; \vec{Q}\cdot \vec{r}-
\vec{Q}\cdot \vec{r} \; \vec{Q}\cdot \vec{\tg} \; \; ,\]
where
\[ \vec{Q}\equiv (i, \; j, \;k) \; \; \; , \; \; \;
\vec{\tg}\equiv (\ag, \; \bg, \; \cg) \; \; .\]
If we rewrite the mentioned above transformation in the following
form\f{{\em Barred} operators ${\cal O}\mid q$ act on quaternionic objects
$\Phi$ as in
\[({\cal O}\mid q)\Phi={\cal O} \Phi q \; \;. \]}
\be
\vec{Q}\cdot \vec{r} \; ' = [1+\vec{\tg}\cdot (\vec{Q} -1 \mid \vec{Q})]
\; \vec{Q}\cdot \vec{r} \; \; ,
\ee
we identify
\[ \frac{i - 1\mid i}{2} \; \; \; , \; \; \;
   \frac{j - 1\mid j}{2} \; \; \; , \; \; \;
   \frac{k - 1\mid k}{2} \; \; \; ,\]
as the generators for rotations in the three-dimensional space\f{The factor
$\frac{1}{2}$ guarantees that our generators satisfy the usual algebra:
\[ [A_{m}, \; A_{n}]=\eg_{mnp} A_{p} \; \; \; \; \; m, \; n, \; p =1, \; 2,
\; 3 \; \; .\]}.

Up till now, we have considered only particular operations on
quaternions. A quaternion $q$ can also be multiplied by unitary
quaternions $\cal V$ from the right. A possible transformation which
preserves the norm is given by
\bel{o4}
q'={\cal U} q \; {\cal V} \; \; ,
\ee
\bc
{\fs $( \; {\cal U}^{+}{\cal U}={\cal V}^{+}{\cal V}=1 \; ) \; \; .$}
\ec
Since left and right multiplications commute, the group is locally isomorphic
to $SU(2)\times SU(2)$, and so to $O(4)$, the four-dimensional Euclidean
rotation group.

As far as here we can recognize only particular real linear quaternions,
namely
\[ 1 \; \; , \; \; i \; \; , \; \; j \; \; , \; \; k \; \; , \; \;
1 \; \vert \; i \; \; , \; \; 1 \; \vert \; j \; \; , \; \;
1 \; \vert \; k \; \; . \]
We haven't hope to describe the Lorentz group if we use only previous
objects. Analyzing the most general transformation on
quaternions (see section 4), we introduce new real linear quaternions which
allow us to overcome the above difficulty and so
obtain a quaternionic version of the Lorentz group, without the use of
complexified quaternions. This result
appears, to the best of our knowledge, for the first time in print.\\
First of that we briefly recall the standard way to rewrite special
relativity by a quaternionic algebra on complex (see section 3).\\
In section 5, we present a quaternionic version of the special group
$SL(2, \; c)$, which is as well-known collected to the Lorentz group.
Our conclusions are drawn in the final section.

\section{Quaternionic algebras}

A quaternionic algebra over a field $\cal F$ is a set
\[ {\cal H} = \{ \ag + i \bg + j \cg + k \dg \; \; \vert
\; \; \ag , \; \bg , \; \cg , \; \dg \; \in \; {\cal F} \} \]
 with multiplication operations defined by following rules for imaginary
units $i, \; j, \; k$
\bean
i^{2}=j^{2}=k^{2} & = & -1 \; \; ,\\
jk=-kj            & = & i \; \; , \\
ki=-ik            & = & j \; \; , \\
ij=-ji            & = & k \; \; .
\eean
In our paper we will work with quaternionic algebras defined on reals and
complex, so in this section we give a panoramic review of such algebras.

We start with a quaternionic algebra on reals
\[ {\cal H}_{\cal R} = \{ \ag + i \bg + j \cg + k \dg \; \; \vert
\; \; \ag , \; \bg , \; \cg , \; \dg \; \in \; {\cal R} \}  \; \; .\]
We introduce the quaternion conjugation denoted by $^{+}$ and defined by
\[ q^{+}=\ag - i\bg -j\cg - k\dg \; \; . \]
The previous definition implies
\[ (\psi \vpg)^{+} = \vpg^{+} \psi^{+} \; \; ,\]
for $\psi$, $\vpg$ quaternionic functions. A conjugation
operation which does not reverse the order of $\psi$, $\vpg$ factors is
given, for example, by
\[ \tilde{q} = \ag - i\bg +j\cg - k\dg \; \; . \]
An important difference between quaternions and complexified
quaternions is based on the concept of {\em division algebra}, which is a
finite dimensional algebra for which $a\neq 0$, $b\neq 0$ implies
$ab\neq 0$, in others words, which has not nonzero divisors of zero. A
classical theorem~\cite{bot} states that the only division algebras over
the reals are algebras of dimension $1$, $2$, $4$ and $8$; the only
associative algebras over the reals are $\cal R$, $\cal C$ and
${\cal H}_{\cal R}$ (Frobenius~\cite{fro}); the nonassociative division
algebras include the octonions $\cal O$ (but there are others as well; see
Okubo~\cite{oku}).\\
A simple example of a {\em nondivision} algebra is provided by the algebra
of complexified quaternions
\[ {\cal H}_{\cal C} = \{ \ag + i \bg + j \cg + k \dg \; \; \vert
\; \; \ag , \; \bg , \; \cg , \; \dg \; \in \;
{\cal C}(1, \; {\cal I}) \}  \; \; ,\]
\[ [{\cal I}, \; i]=[{\cal I}, \; j]=[{\cal I}, \; k]=0 \; \; .\]
In fact since
\[ (1+i{\cal I})(1-i{\cal I})=0 \; \; ,\]
there are nonzero divisors of zero.

For complexified quaternions we have different opportunities to define
conjugation operations, we shall use the following terminology:
\ben
\item The {\sl complex} conjugate of $q_{\cal C}$ is
\[ q_{\cal C}^{\; *} = {\ag}^{*} + i{\bg}^{*} +j{\cg}^{*} +k{\dg}^{*} \; \; .
\]
Under this operation
\bean
({\cal I}, \; i, \; j, \; k) & \rightarrow & (-{\cal I}, \; i, \; j, \; k)
\; \; ,
\eean
and
\[ (q_{\cal C}p_{\cal C})^{*}=q_{\cal C}^{\; *} p_{\cal C}^{\; *} \; \; .\]
\item The {\sl quaternion} conjugate of $q_{\cal C}$ is
\[ q_{\cal C}^{\; \star} = \ag - i\bg -j\cg -k \dg \; \; . \]
Here
\bean
({\cal I}, \; i, \; j, \; k) & \rightarrow & ({\cal I}, \; -i, \; -j, \; -k)
\; \; ,
\eean
and
\[ (q_{\cal C}p_{\cal C})^{\star}=p_{\cal C}^{\; \star}
q_{\cal C}^{\; \star} \; \; .\]
\item In the absence of standard terminology, we call that formed by
combining these operations, the {\sl full} conjugate
\[ q_{\cal C}^{\; +} = {\ag}^{*} - i{\bg}^{*} -j{\cg}^{*} -k{\dg}^{*} \; \; .
\]
Under this operation
\bean
({\cal I}, \; i, \; j, \; k) & \rightarrow & -({\cal I}, \; i, \; j, \; k)
\; \; ,
\eean
and
\[ (q_{\cal C}p_{\cal C})^{+}=p_{\cal C}^{\; +}
q_{\cal C}^{\; +} \; \; .\]
\een

\section{Complexified Quaternions and Special Relati\-vi\-ty}

We begin this section by recalling a sentence of Anderson
and Joshi~\cite{and} about the quaternionic reformulation of special
relativity:

``{\sl There has been a long tradition of using quaternions for Special
Relativity ... The use of quaternions in special relativity, however, is not
entirely straightforward. Since the field of quaternions is a
four-dimensional Euclidean space, complex components for the quaternions
are required for the $3+1$ space-time of special relativity}.''

In the following section, we will demonstrate
that a reformulation of special relativity by a quaternionic algebra on
reals is possible.

In the present section, we use complexified
quaternions to reformulate special relativity, for further details the reader
may consult the papers of Edmonds~\cite{edm}, Gough~\cite{gou},
Abonyi~\cite{abo}, G\"ursey~\cite{gur} and the book of Synge~\cite{syn}.

A space-time point can be represented by complexified quaternions as follows
\be
{\cal X} = {\cal I}ct + ix+jy+kz \; \; .
\ee
The Lorentz invariant in this formalism is given by
\be
{\cal X}^{*}{\cal X}=(ct)^{2} - x^{2} - y^{2} - z^{2} \; \; .
\ee
If we consider the standard Lorentz transformation (boost $ct$ - $x$)
\bean
ct' & = & \gamma \; (ct-\beta x) \; \; ,\\
x' & = & \gamma \; (x-\beta ct) \; \; ,\\
y' & = & y \; \; ,\\
z' & = & z \; \;
\eean
and note that the first two equations may be rewritten as
\bean
ct' & = & ct \cosh{\theta} - x \sinh{\theta} \; \; ,\\
x' & = & x \cosh{\theta} - ct \sinh{\theta} \; \; ,
\eean
\bc
{\fs where $\cosh{\tg} =\cg \; \; \; , \; \; \;
\sinh{\tg} = \bg \cg \; \; ,$}
\ec
we can represent an infinitesimal transformation by
\bean
{\cal X}' & = & {\cal I}(ct-x\theta ) +i(x -ct\theta )+jy+kz \; =\\
          & = & {\cal X} + {\cal I} \; \frac{i+1\mid i}{2} \tg {\cal X} \; \; .
\eean
We thus recognize, in the previous transformation, the generator
\[ {\cal I} \; \frac{i+1\mid i}{2} \; \; . \]
It is now very simple to complete the translation, the set of generators
of the Lorentz group is provided with
\bean
boost \; \; (ct, \; x) \; \; &
\; \;  {\cal I} \; \frac{i+1 \; \vert \; i}{2} \; \; ,\\
boost \; \; (ct, \; y) \; \; &
\; \;  {\cal I} \; \frac{j+1 \; \vert \; j}{2} \; \; ,\\
boost \; \; (ct, \; z) \; \; &
\; \;  {\cal I} \; \frac{k+1 \; \vert \; k}{2} \; \; ,\\
rotation \; \; around \; \; x \; \; &
\; \; \; \; \; \frac{i-1 \; \vert \; i}{2} \; \; ,\\
rotation \; \; around \; \; y \; \; &
\; \; \; \; \; \frac{j-1 \; \vert \; j}{2} \; \; ,\\
rotation \; \; around \; \; z \; \; &
\; \; \; \; \; \frac{k-1 \; \vert \; k}{2} \; \; .
\eean
And so a general finite Lorentz transformation is given by
\[ e^{{\cal I}(i\alpha_{b}+j\beta_{b}+k\gamma_{b})+
i\alpha_{r} +j\beta_{r} +k \gamma_{r}} ({\cal I}ct + ix+jy+kz)
e^{{\cal I}(i\alpha_{b}+j\beta_{b}+k\gamma_{b})-
i\alpha_{r} -j\beta_{r} -k \gamma_{r}} \; \; . \]
The previous results can be elegantly summarize by the relation
\be
{\cal X}'=\Lg {\cal X} \Lg^{+}  \; \; \; \; , \; \; \; \;
\Lg^{\star} \Lg = 1 \; \; ,
\ee
where $\Lg$ is obviously a complexified quaternion. In this or similar way
a lot of authors have reformulated special relativity with complex
quaternions.

In the following section we will introduce {\em real linear quaternions}
and reformulate special relativity using a quaternionic algebra on
reals. We remark that complex component for the quaternions represent a
choice and not a necessity.

\section{A new possibility}

We think that quaternions are the natural candidates to
describe special relativity. It is simple to understand why,
quaternions are characterized by four real numbers (whereas complexified
quaternions by eight), thus we can collect these four
real quantities with a point $(ct, \; x, \; y, \; z)$ in the space-time.
In quaternionic notation we have
\begin{equation}
{\cal X}= ct + ix + jy + kz \; \; .
\end{equation}
In the first section we have introduced particular {\em real linear
quaternions}, namely
\[ 1 \; \; , \; \; \vec{Q} \; \; , \; \;  1\mid  \vec{Q} \; \; ,\]
where
\[ \vec{Q}\equiv (i, \; j, \; k) \; \; . \]
In order to write the most general real linear quaternions we must consider
the following quantities
\[ \vec{Q}\mid i \;  \; , \; \; \vec{Q}\mid j \; \; , \; \; \vec{Q}\mid k
\; \; .\]
In fact the most general transformation on quaternions is represented by
\bel{w}
q+ p\mid i + r\mid j + s\mid k  \; \; ,
\ee
with
\[ q, \; p, \; r, \; s \;
\in \; {\cal H}_{\cal R} \; \; . \]
New objects like
\[ k\mid j \; \; , \; \; j\mid k \; \; , \; \; i\mid k \; \; , \; \;
k\mid i \; \; , \; \; j\mid i \; \; , \; \; i\mid j \]
will be essential to reformulate special relativity with real quaternions.
They represent the wedge which permit to overcome the difficulties which
in past did not allow a (real) quaternionic version of special relativity.

Returning to Lorentz transformations, let us start with the following
infinitesimal transformation  (boost $ct$ - $x$)
\bean
{\cal X}' & = & ct-x\theta  +i(x -ct\theta )+jy+kz \; =\\
       & = & {\cal X} + \frac{k\mid j - j\mid k}{2} \; \tg {\cal X} \; \; .
\eean
We can immediately note that the generator which substitutes
\[ {\cal I} \; \frac{i+1\mid i}{2} \]
is
\[ \frac{k\mid j - j\mid k}{2} \; \; . \]
So we have (for the first time in print) the possibility to list the
generators of the Lorentz group without the need to work with complexified
quaternions
\bean
boost \; \; (ct, \; x) \; \; &
\; \; \frac{k \; \vert \; j - j \; \vert \; k}{2} \; \; ,\\
boost \; \; (ct, \; y) \; \; &
\; \; \frac{i \; \vert \; k - k \; \vert \; i}{2} \; \; ,\\
boost \; \; (ct, \; z) \; \; &
\; \; \frac{j \; \vert \; i - i \; \vert \; j}{2} \; \; ,\\
rotation \; \; around \; \; x \; \; &
\; \; \; \;  \frac{i-1 \; \vert \; i}{2} \; \; ,\\
rotation \; \; around \; \; y \; \; &
\; \; \; \;  \frac{j-1 \; \vert \; j}{2} \; \; ,\\
rotation \; \; around \; \; z \; \; &
\; \; \; \;  \frac{k-1 \; \vert \; k}{2} \; \; .
\eean

In appendix A we explicitly prove that the action of previous generators
leaves
\begin{equation}
Re \; {\cal X}^{2} = (ct)^{2} - x^{2} - y^{2} - z^{2}
\end{equation}
invariant.

In appendix B we will give an alternative but equivalent presentation of
special relativity by a quaternionic algebra on reals. There we introduce a
real linear quaternion $g$ which substitutes the metric tensor $g^{\mu \nu}$.

\section{A quaternionic version of the complex group $SL(2)$}

In analogy to the connection between the rotation group
$O(3)$ to the special unitary group $SU(2)$, there is a natural
correspondence~\cite{tun} between
the Lorentz group $O(3,~1)$ and the special linear group
$SL(2)$. In fact $SL(2)$ is the universal covering group of $O(3,~1)$ in
the same way that $SU(2)$ is of $O(3)$.

The aim of this Section is to give, by extending the consideration which
collect the special unitary group $SU(2)$ with unitary real quaternions
(as shown in section 1), a quaternionic version of
the special linear group $SL(2)$.
Once more the aim will be achieved with help of real linear quaternions.

A Lorentz spinor is a complex object which transform under Lorentz
transformations as
\[ \psi' = {\cal A} \psi \; \; , \]
where $\cal A$ is a $SL(2)$ matrix. When we restrict ourselves to the
three-dimensional space and to rotations, this definition gives the usual
Pauli spinors
\[ \psi' = {\cal U} \psi \; \; , \]
where $\cal U$ is a $SU(2)$ matrix.

Now we shall derive the generators of rotations and Lorentz boosts in the
spinor representation by using real linear quaternions.

The action of generators of the special group $SL(2)$
\bean
\left( \begin{array}{cc} $i$ & 0\\ 0 & $-i$\end{array} \right) \; , &
\left( \begin{array}{cc} 0 & $-1$\\ 1 & 0\end{array} \right) \; , &
\left( \begin{array}{cc} 0 & $-i$\\ $-i$ & 0\end{array} \right) \; ,\\ \\
\left( \begin{array}{cc} $-1$ & 0\\ 0 & 1\end{array} \right) \; , &
\left( \begin{array}{cc} 0 & $-i$\\ $i$ & 0\end{array} \right) \; , &
\left( \begin{array}{cc} 0 & 1\\ 1 & 0\end{array} \right) \; ,
\eean
on the spinor
\[ \psi = \left( \begin{array}{c} \xi \\ \eta \end{array} \right) \; , \]
can be represented by the action of real linear quaternions
\bean
i \; , & j \; , & k \; ,\\
i \; \vert \; i \; , & j \; \vert \; i \; , & k \; \vert \; i \; ,
\eean
on the quaternion
\[ q= \xi +j\eta \; \; . \]

In section 1 we have obtained a three-dimensional vector $(x, \; y, \; z)$
by product of Pauli spinors $q_{\cal P}$:
\[q_{\cal P}\; i \; q_{\cal P}^{+}=ix+jy+kz \]
\bc
{\fs $ ( \; q'_{\cal P}={\cal U}q_{\cal P} \; \; \; , \; \; \;
{\cal U}^{+}{\cal U}=1 \; ) \; \; ,$}
\ec
consequently we have written its transformation law as follows
\[(q_{\cal P} \; i \; q_{\cal P}^{+})'={\cal U}q_{\cal P} \; i \;
q_{\cal P}^{+} \; {\cal U}^{+}
\; \; .\]

Now we start with a Lorentz spinor $q_{\cal L}$
\[q'_{\cal L}={\cal A}q_{\cal L} \; \; ,\]
and construct a four-vector $(ct, \; x, \; y, \; z)$ by product of such
spinors
\[q_{\cal L} \; (1+i) \; q_{\cal L}^{+}=ct+ix+jy+kz \; \; .\]
The transformation law is then given by
\[(q_{\cal L} \; (1+i) \; q_{\cal L}^{+})'=({\cal A}q_{\cal L}) \; (1+i)
\; ({\cal A}q_{\cal L})^{+} \; \; .\]
If we consider infinitesimal transformations
\[ {\cal A} = 1+\frac{\vec{Q}}{2}\cdot (\vec{\tg}+\vec{\zeta}\mid i) \; \; ,\]
\bc
{\fs with $\vec{\tg}\equiv (\ag, \; \bg, \; \cg)$ and
$\vec{\zeta}\equiv (\tilde{\ag}, \; \tilde{\bg}, \; \tilde{\cg}) \; \; ,$}
\ec
we have
\bean
{\cal T}' & = & {\cal T}+\frac{\ag}{2}[i, \; {\cal T}]+
\frac{\bg}{2}[j, \; {\cal T}]+\frac{\cg}{2}[k, \; {\cal T}]+\\
 & & + \; \frac{\tilde{\ag}}{2} \{i, \; \tilde{{\cal T}}\}+
\frac{\tilde{\bg}}{2} \{j, \; \tilde{{\cal T}}\}+
\frac{\tilde{\cg}}{2}\{k, \; \tilde{{\cal T}} \}
\eean
where
\[ {\cal T}= q_{\cal L} \; (1+i) \; q_{\cal L}^{+} \; \; ,\]
and
\[ \tilde{{\cal T}} =q_{\cal L} \; i(1+i) \; q_{\cal L}^{+}=
{\cal T} -2q_{\cal L}q_{\cal L}^{+}\; \; .\]
In order to simplify next considerations we pose
\bean
{\cal T}=ix+jy+kz+ct & = {\cal T}_{i}+{\cal T}_{j}+{\cal T}_{k}+
{\cal T}_{1} \; \; , \\
\tilde{\cal T}=ix+jy+kz-ct & = {\cal T}_{i}+{\cal T}_{j}+{\cal T}_{k}-
{\cal T}_{1} \; \; ,
\eean
so the standard Lorentz transformations are given by
\bean
{\cal T}_{1} & \rightarrow & {\cal T}_{1} +\tilde{\alpha} i {\cal T}_{i}+
\tilde{\beta} j {\cal T}_{j}+\tilde{\gamma} k {\cal T}_{k}\\
{\cal T}_{i} & \rightarrow & {\cal T}_{i} -\tilde{\alpha} i {\cal T}_{1}+
\beta j {\cal T}_{k}-\gamma k {\cal T}_{j}\\
{\cal T}_{j} & \rightarrow & {\cal T}_{j} -\tilde{\beta} j {\cal T}_{1}-
\alpha i {\cal T}_{k}+\gamma k {\cal T}_{i}\\
{\cal T}_{k} & \rightarrow & {\cal T}_{k} -\tilde{\gamma} k {\cal T}_{1}+
\alpha i {\cal T}_{j}-\beta j {\cal T}_{i} \; \; .
\eean

In this way we obtain a quaternionic version of the special
group $SL(2)$ and demonstrate\f{In contrast with the opinion of
Penrose~\cite{pen}, cited in footnote 3.} that, if real linear
quaternions appear, a `trick' similar to that one of rotations works to
relate the full four-vector $(ct, \; x, \; y, \; z)$  with real
quaternions.

\section{Conclusions}

The study of special relativity with a quaternionic algebra on reals has
yielded a result of interest. While we cannot demonstrate in this paper
that one number system (quaternions) is preferable to another
(complexified quaternions) we have pointed out the advantages of using
real linear quaternions which naturally appear when we work with a non
commutative number system, like the quaternionic field. As seen in this paper
these objects are very useful if we wish to rewrite special relativity by a
quaternionic algebra on reals. The complexified
quaternionic reformulation of special relativity is thus a choice and
not a necessity. This affirmation is in contrast with the standard folklore
(see, for example, Anderson and Joshi~\cite{and}).

Our principal aim in this work is to underline the potentialities of real
linear quaternions. We wish to remember that a lot of difficulties have
been overcome thanks to these objects (which in our colourful
language we have named generalized objects~\cite{del}).\\
To remark their potentialities let us list the situations which have
requested their use.
\bii
\item The need of such objects naturally appears, for
example, in the construction of quaternion group theory and tensor
product group representations~\cite{del2}. Also starting with only standard
quaternions $i, \; j, \; k$ in order to represent the generators of
the group $U(1, \; q)$, we find generalized quaternions when we analyze
quaternionic tensor products.
\[ Spin \; \; \frac{1}{2} \; \; generators  \; : \; \;
\frac{i}{2} \; \; ,  \; \; \frac{j}{2} \; \; , \; \; \frac{k}{2} \;  \; .\]
\[ Spin \; \; 1\oplus 0 \; \; generators \; :\]
\[ \left( \begin{array}{cc} \frac{i+1 \; \vert \; i}{2} & 0\\
0 & \frac{i-1 \; \vert \; i}{2} \end{array} \right)  \; ,  \;
\left( \begin{array}{cc} j & 1 \; \vert \; i\\
1 \; \vert \; i & j\end{array} \right)  \; , \;
\left( \begin{array}{cc} k & -1\\
1 & k\end{array} \right)  \; \; .\]
\item If we desire to extend the isomorphism of $SU(2, \; c)$ with
$U(1, \; q)$ to the group $U(2, \; c)$, we must introduce the additional
real linear quaternion `$1 \; \vert \; i$'.
In this way there exists at least one version
of quaternionic quantum mechanics in which a `partial' set of translations
may be defined~\cite{del}, in fact,  tanks to real linear operators, a
translation between $2n\times 2n$ complex  and $n\times n$ quaternionic
matrices is possible.
\item In the work of ref.~\cite{rot} a quaternion version of the Dirac
equation was derived in the form
\[ \gamma_{\mu} \partial^{\mu} \psi i = m \psi \; \; ,\]
where the $\gamma_{\mu}$ are two by two quaternionic matrices satisfying the
Dirac condition
\[ \{\gamma_{\mu}, \; \gamma_{\nu} \} = 2 g_{\mu \nu} \; \; . \]
In the Rotelli's formalism the momentum operator must be defined as
\[ p^{\mu} = \partial^{\mu} \; \vert \; i \]
which is also a generalized object.
\item In this paper, contrary to the common opinion, we have given a real
quaternionic formulation of special relativity. In order to obtain that we have
introduced the following real linear quaternions
\[ \vec{Q} \; \vert \; i \; , \;  \vec{Q} \; \vert \; j \; ,
\; \vec{Q} \; \vert \; k \; \; , \]
\[ \vec{Q} \equiv (i, \; j, \; k) \; \; . \]
A quaternionic version of the special group $SL(2)$ has also been given.
\eii

We finally note that the process of generalization can be extend also
to complexified quaternions. In a recent paper~\cite{del3} we give an
elegant one-component formulation of the Dirac equation and, thanks to our
generalization, we overcome previous
difficulties concerning the doubling of solutions~\cite{and,edm,gou} in the
complexified quaternionic Dirac equation.

In seeking a better understanding of the success of mathematical
abstraction in physics and in particular of the wide applicability of
quaternionic numbers in theories of physical phenomena, we found that
generalized quaternions should be {\em not} undervalued. We think
that there are good reasons to hope that these generalized structures provide
new possibilities concerning physical applications of quaternions.

``{\sl The most powerful method of advance that can be suggested at present
is to employ all the resources of pure mathematics in attempts to perfect
and generalize the mathematical formalism that forms the existing basis of
theoretical physics, and after each success in this direction, to try to
interpret the new mathematical features in terms of physical entities...}''
- Dirac~\cite{dir}.

\section*{Appendix A}

In this appendix we prove that the Lorentz invariant is
\bel{apb}
Re \; {\cal X}'^{\; 2}=Re \; {\cal X}^{2} \; \; ,
\ee
where
\[{\cal X}=ct+ix+jy+kz \; \; .\]
Under an infinitesimal transformation, we have
\[ {\cal X}' = (1+\theta \; \frac{k \; \vert \; j-j \; \vert \; k}{2}+
\alpha \; \frac{i-1 \; \vert \; i}{2}+...){\cal X} \]
so, neglecting second order terms,
\[ {\cal X}'^{\; 2} = {\cal X}^{2}+\frac{\theta}{2} \; \{{\cal X}, \;
k{\cal X}j-j{\cal X}k \}+
\frac{\alpha}{2} \; \{{\cal X}, \; i{\cal X}-{\cal X}i \}+ ... \]
Equation~(\ref{apb}) is then satisfied since
\bean
\{{\cal X}, \; i{\cal X}-{\cal X}i \}   & = & (i-1\mid i) {\cal X}^{2} \; \;
,\\
\{{\cal X}, \; k{\cal X}j-j{\cal X}k \} & = & (1\mid j -j) {\cal X}k{\cal X} +
(k - 1\mid k) {\cal X}j{\cal X} \; \; ,
\eean
are purely imaginary quaternions.

Obviously we can derive the generators of Lorentz group by starting
from the infinitesimal transformation
\[ {\cal X}'={\cal X}+{\cal A}{\cal X} \]
and imposing that they satisfy the
relation
\be
Re \; \{{\cal X}, \; {\cal A}{\cal X} \}=0
\ee
\bc
{\fs $( \; Re \; {\cal X}'^{\; 2}=Re \; {\cal X}^{2} \; \;
\Raw \; \; Re \; \{{\cal X}, \; {\cal A}{\cal X} \}=0 \; )$ .}
\ec
With straightforward mathematical calculus we can find the generators
requested. In order to simplify following considerations let us pose
\[ {\cal X}=a+ib+jc+kd \; \; \; , \; \; \;
{\cal A}=q_{0}+q_{1}\mid i+q_{2}\mid j+q_{3}\mid k\]
\bc
{\fs where $q_{m}={\alpha}_{m}+i{\beta}_{m}+j{\gamma}_{m}+k{\delta}_{m} \; \;
(m \; = \; 0, \; 1, \; 2, \; 3)$ are real quaternions.}
\ec
The only quantities which we must calculate are
\[ Re \; \{{\cal X}, \; {\cal X} \} \; \; , \; \;
Re \; \{{\cal X}, \; i{\cal X}i \} \; \; , \; \;
Re \; \{{\cal X}, \; i{\cal X} \} \; \; , \; \;
Re \; \{{\cal X}, \; k{\cal X}j \} \; \; ,\]
in fact the other quantities can be obtained from previous ones, by simple
manipulations.
\bean
Re \; \{{\cal X}, \; {\cal X} \} \; \; =2(+a^{2}-b^{2}-c^{2}-d^{2}) & , &
Re \; \{{\cal X}, \; i{\cal X}i \} \; =2(-a^{2}+b^{2}-c^{2}-d^{2})\\
Re \; \{{\cal X}, \; j{\cal X}j \}=2(-a^{2}-b^{2}+c^{2}-d^{2}) & , &
Re \; \{{\cal X}, \; k{\cal X}k \}=2(-a^{2}-b^{2}-c^{2}+d^{2})\\
Re \; \{{\cal X}, \; i{\cal X} \}=Re \; \{{\cal X}, \; {\cal X}i \}=-4ab & , &
Re \; \{{\cal X}, \; k{\cal X}j \}=Re \; \{{\cal X}, \; j{\cal X}k \}=4cd\\
Re \; \{{\cal X}, \; j{\cal X} \}=Re \; \{{\cal X}, \; {\cal X}j \}=-4ac & , &
Re \; \{{\cal X}, \; j{\cal X}i \} \; =Re \; \{{\cal X}, \; i{\cal X}j \} \;
=4bc\\
Re \; \{{\cal X}, \; k{\cal X} \}=Re \; \{{\cal X}, \; {\cal X}k \}=-4ad & , &
Re \; \{{\cal X}, \; i{\cal X}k \}\; =Re \; \{{\cal X}, \; k{\cal X}i \} \;
=4bd \; \; .
\eean
The previous relations imply the following conditions on the real parameters
of the generator $\cal A$
\bean
\alpha_{0} = 0 & , & \beta_{1} = 0 \\
\gamma_{2} = 0 & , & \delta_{3} = 0 \\
\beta_{0} = - \alpha_{1} = \alpha & , & \gamma_{0} = - \alpha_{2} = \beta \\
\delta_{0} = - \alpha_{3} = \gamma & , & \delta_{2} = - \gamma_{3} = \theta \\
\gamma_{1} = - \beta_{2} = \varphi & , & \beta_{3} = - \delta_{1} = \eta \; \;
{}.
\eean
We can immediately recognize the Lorentz generators given in section 4.

\section*{Appendix B}

We introduce the usual four-vector $x^{\mu}$ by the following
quaternion
\[ {\cal X} = x^{0} + i x^{1} + j x^{2} + k x^{3} \; \; , \]
and define a scalar product of two vectors $\cal X$, $\cal Y$ by
\begin{equation} \label{rsp}
({\cal X}, \; g{\cal Y})_{\cal R} = Re \; ({\cal X}^{+}g{\cal Y}) =
x^{\mu} g_{\mu \nu} y^{\nu} \; \; ,
\end{equation}
where $g$ is the generalized quaternion
\[ - \frac{1}{2} \; (1 + i \; \vert \; i + j \; \vert \; j +
k \; \vert \; k)  \; \; .\]
We can define a real norm (or metric)
\[ ({\cal X}, \; g{\cal X})_{\cal R} = Re \; ({\cal X}^{+}g{\cal X}) =
x^{\mu} g_{\mu \nu} x^{\nu} \; \; .\]
The vectors which transform under a Lorentz transformation ${\cal L}$, will be
denoted by
\[ {\cal X}' = {\cal L}{\cal X} \; \; , \]
with $\cal L$ real linear operators (see eq.~(\ref{w})).\\ From the
postulated invariance of the norm we can deduce the generators of Lorentz
group.

If we consider infinitesimal transformations
\[ {\cal L} = 1 + {\cal A} \; \; ,\]
we have
\[ Re \; ({\cal X}'^{+}g{\cal X}') = Re \; ({\cal X}^{+}g{\cal X} +
{\cal X}^{+}( {\cal A}^{+}g + g{\cal A}){\cal X}) =
Re \; ({\cal X}^{+}g{\cal X})
\; \; ,\]
and therefore
\bel{gen}
{\cal A}^{+}g + g{\cal A} = 0 \; \; .
\ee
Using real scalar products, given an operator
\[ {\cal A}=q+p \; \vert \; i+r \; \vert \; j+s \; \vert \; k \; \; ,\]
\[ q, \; p, \; r, \; s \; \in \; {\cal H}_{\cal R} \; \; ,\]
we can write its hermitian conjugate as follows
\[ {\cal A}^{+}=q^{+}-p^{+} \; \vert \; i-r^{+} \; \vert \; j-
s^{+} \; \vert \; k \; \; .\]
Then the equation~(\ref{gen}) can be rewritten as
\[ g{\cal A}+h.c.=0 \; \; .\]
If we pose
\[ g{\cal A}=B= \tilde{q}+ \tilde{p} \; \vert \; i+\tilde{r} \; \vert \; j+
\tilde{s} \; \vert \; k \; \; ,\]
we obtain the following conditions on the operator $B$
\[ Re \; \tilde{q} = Vec \; \tilde{p} = Vec \; \tilde{r} =
Vec \; \tilde{s} = 0 \; \; .\]
Noting that ${\cal A}=gB$ we can quickly write the generators of
Lorentz group. We give explicitly an example
\[ {\cal A}_{1}=g \; (1 \; \vert \; i) = -\frac{1}{2}(-i+1 \; \vert \; i +
j \; \vert \; k - k \; \vert \; j) \; \; , \]
\[ {\cal A}_{2}=g \; i = -\frac{1}{2}(i-1 \; \vert \; i +j \; \vert \; k -
k \; \vert \; j) \; \; ,\]
\bean
{\cal A}={\cal A}_{1}-{\cal A}_{2} & = & \frac{i - 1 \; \vert \; i}{2} \; \;
,\\
\tilde{{\cal A}}={\cal A}_{1}+{\cal A}_{2} & = &
\frac{k \; \vert \; j - j \; \vert \; k}{2} \; \; .
\eean

\bb
\bi{fey}
{\em Feynman RP} - {\sl The Feynman Lectures on Physics}, vol.~I - part 1.\\
Inter European Editions, Amsterdam (1975).
\bi{pap}
{\em Finkelstein D, Jauch JM, Schiminovich S, Speiser D} - \\
J.~Math.~Phys. {\bf 3}, 207 (1962); {\bf 4}, 788 (1963).\\
{\em Finkelstein D, Jauch JM, Speiser D} - J.~Math.~Phys. {\bf 4}, 136
(1963).\\
{\em Adler SL} - Phys.~Rev. {\bf D21}, 550 (1980), {\bf D21}, 2903 (1980);\\
{\em Adler SL} - Phys.~Rev.~Letts {\bf 55}, 783 (1985);\\
{\em Adler SL} - Phys.~Rev. {\bf D34}, 1871 (1986);\\
{\em Adler SL} - Comm.~Math.~Phys. {\bf 104}, 611 (1986);\\
{\em Adler SL} - Phys.~Rev. {\bf D37}, 3654 (1988);\\
{\em Adler SL} - Nuc.~Phys. {\bf B415}, 195 (1994).\\
\pr{De Leo S, Rotelli P}{D45}{575}{92};\\
{\em De Leo S, Rotelli P} - {\sl Quaternions Higgs and Electroweak Gauge
Group} - {\fs to appear in {\bf Int.~J.~of Mod.~Phys.~A}}.\\
{\em De Leo S} - {\sl Quaternions for GUTs} - {\fs submitted for
publication in {\bf J.~Math.~Phys.}}\\
{\em Dimitir\'c R, Goldsmith B} - Math.~Intell. {\bf 11}, 29 (1989).\\
{\em Horwitz LP,  Biedenharn LC} - Ann.~of Phys. {\bf 157}, 432 (1984).\\
{\em Horwitz LP} -J.~Math.~Phys. {\bf 34}, 3405 (1993).\\
{\em Razon A, Horwitz LP} - Acta Appl.~Math. {\bf 24}, 141 (1991); 179
(1991);\\
{\em Razon A, Horwitz LP} - J.~Math.~Phys. {\bf 33}, 3098 (1992).\\
{\em Rembieli\'nski J} - J.~Phys. {\bf A11}, 2323 (1978),
{\bf A13}, 15 (1980);\\
{\em Rembieli\'nski J} - J.~Phys. {\bf A13}, 23 (1980),
{\bf A14}, 2609 (1981).\\
{\em Nash CC and Joshi GC} - J.~Math.~Phys. {\bf 28}, 2883, 2886 (1987);\\
{\em Nash CC and Joshi GC} - Int.~J.~Theor.~Phys. {\bf 27}, 409 (1988);\\
{\em Nash CC and Joshi GC} - Int.~J.~Theor.~Phys. {\bf 31}, 965 (1992).
\bi{boo}
{\em Finkelstein D, Jauch JM, Speiser D} - {\sl Notes on quaternion quantum
mechanics}\\
{\fs in Logico-Algebraic Approach to Quantum Mechanics II}, Hooker (Reidel,
Dordrecht 1979), 367-421.\\
{\em Hamilton WR} - {\sl Elements of Quaternions} - Chelsea Publishing Co.,
N.Y., 1969.\\
{\em Gilmore R} - {\sl Lie Groups, Lie Algebras and Some of their
Applications} - John Wiley \& Sons, 1974.\\
{\em Altmann SL} - {\sl Rotations, Quaternions, and Double Groups} - Claredon,
Oxford, 1986.\\
{\em Adler SL} - {\sl Quaternion quantum mechanics and quantum fields} -
Oxford UP, Oxford, 1995.
\bi{del}
{\em De Leo S, Rotelli P} - Prog.~Theor.~Phys. {\bf 92}, 917 (1994).
\bi{del2}
\nc{De Leo S, Rotelli P}{B110}{33}{95}.
\bi{pen}
{\em Penrose R, Rindler W} - {\sl Spinors and space-time} - Cambridge UP,
Cambridge, 1984 ( vol.~1, pag.~23).
\bi{bot}
{\em Bott R, Milnor J} - Bull.~Amer.~Math.~Soc. {\bf 64}, 87 (1958).\\
{\em Kervaire M} - Proc.~Nat.~Acad.~Sci. {\bf 44}, 280 (1958).
\bi{fro}
{\em Frobenius G} - J.~Reine Angew.~Nath. {\bf 84}, 59 (1878).
\bi{oku}
{\em Okubo S} - {\sl Introduction to Octonion and Other Non-Associative
Algebras in Physics} - {\fs unpublished}.
\bi{and}
{\em Anderson R, Joshi GC} - Phys.~Essays {\bf 6}, 308 (1993).
\bi{edm}
{\em Edmonds JD} - Int.~J.~Theor.~Phys. {\bf 6}, 205 (1972).\\
{\em Edmonds JD} - Found.~of Phys. {\bf 3}, 313 (1973).\\
{\em Edmonds JD} - Am.~J.~Phys. {\bf 42}, 220 (1974).
\bi{gou}
\ejp{Gough W}{7}{35}{86}; \xx{8}{164}{87}; \xx{10}{188}{89}.
\bi{abo}
{\em Abonyi I, Bito JF, Tar JK} - J.~Phys.~A {\bf 24}, 3245 (1991).
\bi{gur}
{\em G\"ursey F} - {\sl Symmetries in Physics (1600-1980): Proceedings of the
$1^{st}$ International Meeting on the History of Scientific Ideas} -
Seminari d'~Historia de les Ciences, Barcellona, Spain,  557 (1983).
\bi{syn}
{\em Synge JL} - {\sl Quaternions, Lorentz Transformation, and the Conway
Dirac Eddington Matrices} - Dublin Institute for Advanced Studies, Dublin,
1972.
\bi{tun}
{\em Wu-Ki Tung} - {\sl Group Theory in Physics} - World Scientific,
Singapore, 1985.
\bi{rot}
{\em Rotelli P} - Mod.~Phys.~Lett.~A {\bf 4}, 933 (1989).
\bi{del3}
{\em De Leo S} - {\sl One-component Dirac equation} - {\fs submitted for
publication in {\bf Int.~J.~of Phys.~A}.}
\bi{dir}
{\em Dirac PAM} - Proc.~R.~Soc.~London {\bf A133}, 60 (1931).

\eb

\end{document}